\documentstyle[preprint,eqsecnum,aps]{revtex}

\begin{document}

\draft

\title{Transformations of units and world's geometry}

\author{ Israel Quiros\thanks{israel@uclv.etecsa.cu}}
\address{ Departamento de Fisica. Universidad Central de Las Villas. Santa Clara. CP: 54830 Villa Clara. Cuba }

\date{\today}

\maketitle

\begin{abstract}

The issue of the transformations of units is treated, mainly, in a geometrical context. It is shown that Weyl-integrable geometry is a consistent framework for the formulation of the gravitational laws since the basic law on which this geometry rests is invariant under point-dependent transformations of units. Riemann geometry does not fulfill this requirement. Spacetime singularities are then shown to be a consequence of a wrong choice of the geometrical formulation of the laws of gravitation. This result is discussed, in particular, for the Schwrazschild black hole and for Friedmann-Robertson-Walker cosmology. Arguments are given that point at Weyl-integrable geometry as a geometry implicitly containing the quantum effects of matter. The notion of geometrical relativity is presented. This notion may represent a natural extension of general relativity to include invariance under the group of units transformations. 

\end{abstract}


\section{Introduction}

Seemingly Dicke was the first physicist who called attention upon the importance of the transformations of units in physics\cite{dk}. Under a units transformation the coordinate system is held fixed. Hence, the labeling of the spacetime coincidences is invariant, while the curvature scalar and other purely geometrical scalars, invariant under general coordinate transformations, are generally not invariant under a units transformation. Therefore, spacetime measurements [observations] being nothing but just verifications of the spacetime coincidences, are invariant too under general transformations of units. Moreover, it is evident that the particular values of the units of measure employed are arbitrary, i.e., the physical laws must be invariant under a transformation of units\cite{dk}. This simple argument suggests that Einstein's general relativity(GR) in its actual form due to the action $S_{GR}=\int d^4 x \sqrt{-g}(R+16\pi L_{matter})$, where $R$ is the curvature scalar and $L_{matter}$ is the Lagrangian for the matter fields, is not a complete theory of spacetime. In fact, since the scalar $R$ changes under a general transformation of units, then the laws of gravitation derived from the canonical action for GR change too. This conclusion is not a new one. Einstein's GR is intrinsically linked with the occurrence of spacetime singularities and, it is the hope that, when quantum effects would be included, the singularities would be removed from the description of the physical world. Other arguments showing that canonical Einstein's GR is incomplete come from string theory. This theory suggests that a scalar field (the dilaton) should be coupled to gravity in the low-energy limit of the theory\cite{gsw}.

Brans-Dicke (BD) theory of gravitation [and scalar-tensor(ST) theories in general] represents a natural generalization of Einstein's GR. This theory was first presented in reference \cite{bdk} and subsequently reformulated by Dicke\cite{dk} in a conformal frame where the BD gravitational laws seemed like the Einstein's laws of gravitation. Both formulations of BD theory are linked by the conformal rescaling of the spacetime metric,

\begin{equation}
\hat g_{ab}=\Omega^2(x) g_{ab},
\end{equation}
where $\Omega(x)$ is a smooth, nonvanishing function on the spacetime manifold. Under (1.1) the coordinate system is held fixed as we have already remarked. This transformation can be viewed as a particular transformation of units: a point-dependent scale factor applied to the units of length, time and reciprocal mass\cite{dk}. In effect, if one takes the arc-length as one's unit of measurement [length and time unit since the speed of light $c=1$] hence, since $d\hat s=\Omega(x) ds$, the units of length and time in the conformal frame are point-dependent even if $ds$ is a constant along a geodesic in the original frame. In the same way, in Brans-Dicke theory the mass unit transforms [under (1.1)] as $\hat m=\Omega^{-1}(x)m$. Hence, in the conformal frame the unit of mass $\hat m$ will be point-dependent even if $m$ [the unit of mass in the original formulation] is a constant.

The original formulation of BD theory was based on the following action\cite{bdk}:

\begin{equation}
S_{BD}=\int d^4 x \sqrt{-g}(\phi R-\frac{\omega}{\phi}(\nabla\phi)^2+16\pi L_{matter}),
\end{equation}
where $R$ is the Ricci scalar, $\phi$ is the BD scalar field, and $\omega$ is the BD coupling constant -a free parameter of the theory. Under (1.1) with $\Omega^2=\phi$, (1.2) is mapped into the conformal formulation of BD theory\cite{dk},

\begin{equation}
S_{BD}=\int d^4 x \sqrt{-\hat g}(\hat R-(\omega+\frac{3}{2})(\hat\nabla\psi)^2+16\pi e^{-2\psi} L_{matter}),
\end{equation}
where the scalar field has been redefined according to $\phi\rightarrow e^\psi$. The gravitational laws derived from (1.2) change too under (1.1) with $\Omega^2=\phi$. It is evident from the different forms of the actions (1.2) and (1.3). At the same time, under this transformation, the laws of Riemann geometry change and, what appears to be a Riemann manifold with metric $g_{ab}$ is transformed under (1.1) into a manifold of conformally-Riemannian structure with metric $\hat g_{ab}$\cite{novello,melnikov,iq,qbc}.\footnotemark\footnotetext{The fact that the arbitrariness in the metric tensor (due to the arbitrariness in the choice of the units of measure) raises questions about the significance of Riemann geometry in relativity, was advanced by Brans and Dicke in reference \cite{bdk}} Conformally-Riemannian manifolds are also aknowledged as Weyl-integrable spacetimes (WISTs)\cite{novello}. They are a special kind of Weyl spacetimes that are of particular interest since they are free of the "second clock effect" that leads to observational inconsistencies\cite{vp}. 

Weyl geometry\cite{weyl} is a class of geometries that admit units of measure that may vary along transport. Put in other words; this geometry has a conformal degree of freedom that is related with the freedom in the choice of the units of measure. In this sense it represents a generalization of Einstein's GR. This feature of Weyl geometry in general [and Weyl-integrable geometry (WIG) in particular] is compatible with the fact, noted by Dicke\cite{dk}, that "... there may be more than one feasible way of stablishing the equality of units at different spacetime points. It is evident then, that the equations of motion of matter must be invariant under a general coordinate-dependent transformation of units...".

In Ref. \cite{iqr}, following the spirit of the idea put forth by Dicke in Ref. \cite{dk} and developed in Ref. \cite{dirac,canuto}, I raised the requirement that the laws of physics must be invariant not only under general coordinate transformations but, also, under point-dependent transformations of the units of length, time and mass\footnotemark\footnotetext{These transformations are meaningless from the physical point of view} to a cathegory of a postulate that I called therein as "Brans-Dicke postulate": {\it The laws of physics, including the laws of gravitation, must be invariant under the transformations of the group of point-dependent transformations of the units of length, time and mass}. In the present paper I shall study the consequences this postulate leads to when dealing with the geometrical structure of spacetime. In this sense I shall extend the "Brans-Dicke postulate"\cite{iqr} to our geometrical discussion since, it is evident that, the particular choice of the units of the geometry should not influence the geometrical laws. {\it Any consistent geometrical description should be insensible to the units one chooses}. 

Since there are decisive arguments against the assumption of a general conformal symmetry of the physical laws\cite{cgno},\footnotemark\footnotetext{The elegance and usefulness of this symmetry has been used, for instance, in Ref.\cite{dirac,canuto}} hence the usefulness and scope of the BD postulate should be discussed in detail. In this sense it will be shown that although, in the general case, conformal symmetry is not a symmetry of the laws of gravity, there is a particular set of one-parameter conformal transformations of the metric tensor that constitute a group of symmetry of the conformal formulation of general relativity (conformal GR) [see below] and, consequently, of conformally-Riemannian [Weyl-integrable] geometry. Therefore this group is identified with the one-parameter group of point-dependent transformations of the units of length, time and mass. In consequence, symmetry arguments hint at conformal GR as the most viable [physically complete] formulation of the low-energy laws of gravity [among those studied in the present paper] and at spacetimes of WIST configuration as corresponding underlying manifolds where to put these laws.

The class of conformal rescalings that are usually used to "jump" from one formulation of generic scalar-tensor, non-linear, and Kaluza-Klein theories of gravity [among others] to their conformal formulations can not be interpreted properly as transformations of units. In this sense I shall note that the geometrical interpretation of the conformal transformation (1.1) as a transformation of the units of length, time and mass is not justified in the general case since, in view of the discussion above, invariance under the [physically meaningless] transformations of the units of measure is an obvious requirement any consistently formulated physical law must share. For this reason, in what follows, the conformal transformations that constitute a group of symmetry of conformal GR [and, consequently, of WIG] I shall call as "group of transformations of the units of length, time and mass" or, simply, "transformations of units". Meanwhile, the transformations that allow jumping from the original fromulation of the theory to its conformal formulation I shall call as "proper conformal transformations" or, simply, "conformal transformations"\cite{iqr}. Hence, the set of conformal rescalings of the metric [Eq.(1.1)] consists of the "transformations of units" and of the "conformal transformations". In virtue of the discussion above, this distinction is absolutely neccessary. There are now clear the differences between the statements such as, for instance, "the laws of gravity are not invariant under the conformal rescalings of the metric" and "the laws of gravity must be invariant under the transformations of the units of measure". Both statements are correct. They are not alternative, but, complementary to each other. 

This paper has been motivated, in part, by the ambitious program Brans and Dicke outlined in the classical paper Ref. \cite{bdk}, where they made evident their hope that "...the physical content of the theory should be contained in the invariants of the group of position-dependent transformations of units and coordiante transformations". In part it has been motivated by some interesting features of Weyl-integrable geometry. In particular the possible connection between Weyl-integrable structures and quantum behaviour\cite{novello,london,rankin,dzhunu}. This posibility is briefly investigated in the present paper.

The paper has been organized in the following way. In Sec. II a brief survey on Weyl-integrable geometry is given and the consequences of the linkage of WIST geometry with an effective theory of spacetime are outlined. Sec. III is the main part of this paper. It is fully devoted to the subject of conformal transformations of the kind (1.1). It will be shown, in particular, that the conformal transformation used to go from one formulation of BD and GR theories into their conformal formulations (transformation (1.1) with $\Omega^2=\phi=e^\psi$) is not properly a units transformation. Consequently I define what I call the group of units transformations. In this section we show also, that Weyl-integrable geometry is a consistent framework where to formulate the gravitational laws since the basic requirement on which WIST geometry rests is invariant in respect to the transformations of this group as required by the Brans-Dicke postulate. Riemann geometry does not share this invariance. Hence, the conformal formulation of Einstein's GR in which the underlying manifold is of WIST structure, gives a consistent formulation of the laws of gravity unlike BD theory and the canonical formulation of GR. The issue of the spacetime singularities will be treated in Sec. IV. It will be shown there that singularities that occur in Riemannian spacetimes are removed (not apparently but effectively) in their conformally-Riemannian spacetimes that is made evident, as illustration, for the Schwarzschild metric and for Friedmann-Robertson-Walker cosmology.  In Sec. V  we present some considerations in favour of Weyl-integrable geometry as an intrinsically quantum geometry. Part of these considerations are based in the de Broglie-Bohm quantum theory of motion\cite{bohm} and on an idea advanced in reference \cite{shojai}. Finally, in Sec. VI  we discuss on the meaning of the Brans-Dicke postulate for geometry. We are led to the notion of geometrical relativity that, we hope, will represent a natural extension of general relativity to include invariance under the group of units transformations.

\section{Weyl-integrable geometry and matter couplings}

In Ref. \cite{novello} it has been remarked that both from {\it a priori} and {\it a posteriori} standpoints, in the canonical Einstein's formulation of general relativity that is generated by the action $S_{GR}=\int d^4 x\sqrt{-g}R$, "...the structure of physical spacetime must correspond unequivocally to that of a Riemann manifold...". The same is true for GR with matter content. In general, theories with minimal coupling of the matter content to the metric field are naturally linked with manifolds of Riemann structure\cite{iqr}. In fact, in theories with the matter part of the action of the kind

\begin{equation}
S_{matter}=16\pi\int d^4x\sqrt{-g}\;L_{matter},
\end{equation}
the [time-like] matter particles follow free-motion paths that are solutions of the following differential equation 

\begin{equation}
\frac{d^2 x^a}{ds^2}+\{^{\;\;a}_{mn}\}\frac{dx^m}{ds}\frac{dx^n}{ds}=0,
\end{equation}
where $\{^{\;a}_{bc}\}\equiv\frac{1}{2}g^{an}(g_{bn,c}+g_{cn,b}-g_{bc,n})$ are the Christoffel symbols of the metric. Eq.(2.2) coincides with the equation defining [time-like] geodesic curves in Riemannian spacetimes. In general a Riemann configuration is characterized  by the requirement that the covariant derivatives of the metric tensor vanish, i.e.,

\begin{equation}
g_{ab;c}=0,
\end{equation}
where semicolon denotes covariant differentiation in a general affine sense.\footnotemark\footnotetext{We use, mainly, the notation of Ref.\cite{novello}} Fulfillment of this condition leads the manifold affine connections $\Gamma^a_{bc}$ to become identical to the Christoffel 
symbols $\{^{\;a}_{bc}\}$ of the metric (i.e., $\Gamma^a_{bc}\equiv\{^{\;a}_{bc}\}$). Hence, a Riemann configuration of spacetime is characterized by the  requirement that $g_{ab\|c}=0$, where the double bar denotes covariant differentiation defined through the Christoffel symbols of the metric [instead of the affine connections $\Gamma^a_{bc}$]. This requirement implies that vector lengths do not change under parallel transport, meaning that the units of measure of the geometry are point-independent.

Therefore, in both Jordan frame BD theory that is derivable from the action Eq. (1.2) and Einstein frame general relativity with an additional scalar field that is derivable from

\begin{equation}
S_{GR}=\int d^4 x\sqrt{-g}(R-\alpha(\nabla\psi)^2+16\pi L_{matter}),
\end{equation}
where $\alpha$($\alpha\geq 0$) is a free parameter and $L_{matter}$ is the Lagrangian for the ordinary matter fields,\footnotemark\footnotetext{For $\alpha=0\;(\psi$ arbitrary) or $\psi=const$ we recover standard Einstein's general relativity} the structure of the underlying manifold is Riemannian in nature since both effective actions show minimal coupling of the Lagrangian of the ordinary matter fields to the dilaton. Both theories are, therefore, compatible with a system of point-independent physical units.

Under the conformal rescaling (1.1) with $\Omega^2=\phi=e^\psi$ the actions Eq. (1.2) and Eq. (2.4) are mapped into their conformal actions Eq. (1.3) and 

\begin{equation}
S_{GR}=\int d^4 x \sqrt{-\hat g} e^{-\psi}(\hat R-(\alpha-\frac{3}{2})(\hat\nabla\psi)^2+16\pi e^{-\psi} L_{matter}),
\end{equation}
respectively. The theory that is derivable from this action I shall call as "conformal GR" or, alternatively, "string-frame GR". 

At the same time, under (1.1), manifolds of Riemann structure are mapped onto manifolds of conformally-Riemannian structure also acknowledged as Weyl-integrable spaces(WIST)\cite{novello}. In effect, under the rescaling (1.1) with $\Omega^2=\phi=e^\psi$, the Riemannian requirement Eq. (2.3) is transformed into the following requirement

\begin{equation}
\hat g_{ab;c}=\psi_{,c}\;\hat g_{ab},
\end{equation}
where now semicolon denotes covariant differentiation in a general affine sense with $\hat\Gamma^a_{bc}$ being the [Weyl] affine connection of the conformal manifold. It is given through the Christoffel symbols of the metric with a hat $\{^{\;a}_{bc}\}_{hat}$ and the derivatives of the scalar [dilaton] function $\psi$ as

\begin{equation}
\hat\Gamma^a_{bc}\equiv\{^{\;a}_{bc}\}_{hat}-\frac{1}{2}(\psi_{,b}\;\delta^a_c+\psi_{,c}\;\delta^a_b-\hat g_{bc}\hat g^{an}\;\psi_{,n}).
\end{equation}

If one compares Eq. (2.6) with the requirement of nonvanishing covariant derivative of the metric tensor $\hat g_{ab}$ in the most generic cases of Weyl geometries\cite{novello,weyl}:

\begin{equation}
\hat g_{ab;c}=f_c\;\hat g_{ab},
\end{equation}
in which $f_a(x)$ is the Weyl gauge vector, one arrives at the conclusion that the conformally-Riemannian geometry -characterized by Eq. (2.6)- is a particular type of Weyl geometry in which the gauge vector is the gradient of a scalar function $\psi$ [the dilaton]. 

Therefore, under the transformation (1.1) with $\Omega^2=\phi=e^\psi$, Riemann geometry is mapped into a WIST geometry\cite{novello}. Hence, theories that are derivable from the actions Eq. (1.3) and Eq. (2.5) [that are conformal to Eq. (1.2) and Eq. (2.4) respectively] are naturally linked with manifolds of WIST configuration. In other words, in theories showing non-minimal coupling of the matter fields to the metric, in particular with the matter part of the action of the kind

\begin{equation}
S_{matter}=16\pi\int d^4x\sqrt{-\hat g}\;e^{-2\psi}\;L_{matter}
\end{equation}
[it is conformal to (2.1)], the nature of the underlying manifold is that of a WIST configuration. In this special case of Weyl spaces, under parallel transport, length variations [that are given through $dl=l\;dx^n\psi_{,n}$, where $l\equiv\hat g_{nm} V^n V^m$ is the length of the vector $V^a(x)$ being parallelly transported] are integrable along closed paths: $\oint dl=0$. For this reason, in manifolds of WIST configuration, the disagreement with observations due to the "second clock effect"\footnotemark\footnotetext{The occurrence of additional loss of synchronization due to this effect was the main objection Einstein raised against Weyl's unification theory.}\cite{vp} -that is inherent to Weyl spacetimes in general- is overcome. 

The equations of free-motion of a material test particle that are derivable from Eq. (2.9)

\begin{equation}
\frac{d^2 x^a}{d\hat s^2}+\{^{\;\;a}_{mn}\}_{hat}\frac{dx^m}{d\hat s}\frac{dx^n}{d\hat s}-\frac{\psi_{,n}}{2}(\frac{dx^n}{d\hat s}\frac{dx^a}{d\hat s}-\hat g^{na})=0,
\end{equation}
are conformal to Eq. (2.2). Equations (2.10) coincide with the equations defining geodesic curves in spacetimes of WIST structure. These can also be obtained with the help of the variational principle $\delta\int e^{-\frac{\psi}{2}} d\hat s=0$, that is conformal to $\delta\int ds=0$.

The requirement of nonvanishing covariant derivative of the metric tensor $\hat g_{ab}$ [Eq. (2.6)] implies that vector lengths may vary along transport or, in other words, that the units of measure may change locally. Therefore WIST geometry represents a generalization of Riemann geometry to include units of measure with point-dependent length.

Summing up. Under the conformal rescaling (1.1) with $\Omega^2=\phi=e^\psi$ theories with minimal coupling of the matter fields to the metric [for instance those with the matter part of the action of the kind (2.1)] being uniquely linked with manifolds of Riemann structure, are mapped into theories with non-minimal coupling of the matter fields to the metric [with the matter part of the action of the kind (2.9)] in which the nature of the underlying manifold is of WIST structure. In particular, the Riemannian geodesics of the metric $g_{ab}$ are mapped under (1.1) with $\Omega^2=\phi=e^\psi$ onto geodesics in manifolds of a WIST structure that are specified by the conformal metric $\hat g_{ab}$ and the gauge vector $\psi_{,a}$.{\it Hence in conformal GR, in which the underlying manifold is of WIST structure, the physical interpretation of the experimental observations is to be given on the grounds of the metric with a hat} $\hat g_{ab}$.

Finally we shall remark that neither Riemann geometry nor its conformally-Riemannian [WIST] geometry is preferred on observational grounds since, they are related through the conformal transformation (1.1) that preserves unchanged the labeling of the spacetime coincidence between two particles\cite{dk}. Therefore, experimental observations being nothing but just verifications of these spacetime coincidences, are unchanged under (1.1). This means that experimental measurements are unable to differentiate these geometries. Nevertheless, as we shall see below, this experimental "duality" of the geometrical interpretation of the laws of gravity\cite{iq}, does not mean that these conformal geometrical interpretations are physically equivalent. This will be shown based on considerations of symmetry.

\section{Weyl-integrable geometry and transformation of units}

As we pointed out in Sec. I, Dicke was the first physicist who noticed the importance of the transformations of units in gravitation theory\cite{dk}. He studied, in particular, a units transformation of the kind (1.1), i. e., a point-dependent scale factor applied to the units of length, time and reciprocal mass. Dicke used the transformation (1.1) with $\Omega^2=\phi=e^\psi$ to formulate the BD theory that, in the Jordan frame is given by the action (1.2), in the Einstein frame [action (1.3)].

In this section we shall extend the Brans-Dicke postulate\cite{iqr} to our geometrical discussion since, it is evident that, the particular choice of the units of the geometry should not influence the geometrical laws. {\it Any consistent geometrical description should be insensible to the units one chooses}. 

However one should be very careful since astrophysical observations put forth evidence against the assumption of a general conformal symmetry of the physical laws\cite{canuto}. Simultaneous holding of the Brans-Dicke postulate and the lack of general conformal symmetry of the physical laws seems to be contradictory. For this reason distinction should be made between general conformal symmetry and invariance under transformations of units.

For the purposes of our discussion we shall take the following conformal transformation

\begin{equation}
\bar g_{ab}=e^{\sigma\psi} g_{ab},
\end{equation}
and the scalar function redefinition

\begin{equation}
\bar\psi=(1-\sigma)\psi ,
\end{equation}
where $\sigma$ is some constant parameter. This transformation was introduced in Ref. \cite{far} with a different definition of the scalar factor and scalar field variable. The transformation (1.1) with $\Omega^2=e^\psi$ is a particular case of (3.1) when $\sigma=1$. 

If $\sigma\neq 1$ the transformation (3.1), (3.2) constitutes a one-parameter Abelian group. In fact, a composition of two successive transformations with parameters $\sigma_1\neq 1$ and $\sigma_2\neq 1$ yields a transformation of the same kind with parameter $\sigma_3=\sigma_1+\sigma_2-\sigma_1 \sigma_2\neq 1$, such that $\sigma_3(\sigma_1, \sigma_2)=\sigma_3(\sigma_2, \sigma_1)$, i. e., the group is commutative. The identity of this group is the transformation with $\sigma=0$. The inverse is the transformation with $\bar\sigma=\frac{\sigma}{\sigma-1}$. We see that for $\sigma=1$ the inverse does not exist. Hence, the transformation (1.1) with $\Omega^2=e^\psi$ does not belong to this group.

Under the one-parameter set of transformations (3.1) and (3.2), the basic requirement of a WIST geometry [Eq. (2.6)] is preserved, i.e., under these transformations, Eq. (2.6) is invariant in form [Eq. (4.1) should be previously written as $\hat g_{ab}=e^{\sigma\psi}\bar g_{ab}$]: $\bar g_{ab;c}=\bar\psi\;\bar g_{ab}$. In particular the geodesic equation of WIST geometry [Eq. (2.10)] is invariant in form too under these transformations. Unfortunately it is not true for Riemann geometry; neither the basic requirement that $g_{ab\|c}=0$ nor the geodesic equation (2.2) is invariant under (3.1) and (3.2).

Arguments [of both axiomatic and observational character] have been put foth in the literature that point at WIST geometries as the most viable framework for the geometrical interpretation of the physical laws and, consequently, for the confrontation of physical observations\cite{vp,ja}. Hence, I feel, it is not casual that just WIST geometries possess the above group of symmetry. It is just one of the reasons why in Ref. \cite{iqr} I identified this group of symmetry of Weyl-integrable geometry with the one-parameter group of point-dependent transformations of the units of measure of this geometry. In what follows I shall call the transformations (3.1), (3.2) with $\sigma\neq 1$ simply as "group of units transformations". It is important to remark that the transformation (1.1) with $\Omega^2=\phi=e^\psi$ [it corresponds to the particular case when in (3.1) $\sigma=1$] does not belong to this group and, therefore, it may not be interpreted properly as a transformation of units.

What happens when we approach an effective theory of spacetime?. Take, for instance, the Brans-Dicke theory given by the string frame action [action (2.2) with the replacement $\phi=e^\psi$]

\begin{equation}
S_{BD}=\int d^4 x \sqrt{-g} e^{\psi}(R-\omega(\nabla\psi)^2+16\pi e^{-\psi} L_{matter}).
\end{equation}

It can be verified that (3.3) is invariant in form under (3.1) and (3.2) with $\sigma\neq 1$ only for pure BD or for BD with ordinary matter content with a trace-free stress-energy tensor\cite{far}\footnotemark\footnotetext{For testing BD theory, transformations (3.1) and (3.2) should be completed with a transformation of the BD coupling constant $\bar\omega=\frac{\omega-\frac{3}{2}\sigma(\sigma-2)}{(1-\sigma)^2}$.}. On the other hand, Brans-Dicke theory is naturally linked with Riemann geometry that is not invariant under the transformations of the "group of units transformations". This means that, in view of the BD postulate, BD theory can not be considered as a consistent theory of spacetime. The Einstein frame action for BD theory is not invariant in form under this units transformation even for pure gravity so, it is not a consistent formulation of the laws of gravity too. The same is true for the canonical Einstein's general relativity due to the action (2.4). Moreover, this formulation of general relativity is linked with Riemann geometry that, as we have just remarked, is not invariant in form under (3.1) and (3.2) with $\sigma\neq 1$. This means that canonical Einstein's GR is not yet a complete theory of spacetime. This conclusion is not a new one. It is well-known that canonical general relativity should be completed with quantum effects. It is the hope that, when such a quantum theory of gravity will be worked out, it should be invariant under units transformations of the kind (3.1), (3.2) with $\sigma\neq 1$.

On the contrary the action (2.5) for conformal general relativity is invariant in form under (3.1) and (3.2) with $\sigma\neq 1$, together with the parameter transformation

\begin{equation}
\bar\alpha=\frac{\alpha}{(1-\sigma)^2}.
\end{equation}

For $\sigma=2$ the transformation (3.1), (3.2), (3.4) ($\bar g_{ab}=e^{-2\psi}\hat g_{ab}, \bar\psi=-\psi$) does not touch the free parameter of the theory $\alpha$. Besides, since the "string coupling" ${\it g}\sim e^\psi$, in this case "strong" and "weak" couplings are interchanged [strong-weak symmetry].

The invariance of the pure gravitational part of (2.5) under (3.1), (3.2) and (3.4) with $\sigma\neq 1$ is straightforward. It is essentially the same as the pure gravitational part of BD theory, which invariance under these transformations [with $\alpha=\omega+\frac{3}{2}$] has been checked, for instance, in \cite{far}. For the matter part of (2.5) [it coincides with Eq. (2.9)] we have that, under (3.1) [recall that Eq. (3.1) should be previously written as $\hat g_{ab}=e^{\sigma\psi}\bar g_{ab}$]

\begin{equation}
S_{matter}=16\pi\int d^4 x\sqrt{-\hat g}e^{-2\psi}L_{matter}=16\pi\int d^4 x\sqrt{-\bar g}e^{-2(1-\sigma)\psi}L_{matter},
\end{equation}
since $\sqrt{-\hat g}=e^{2\sigma\psi}\sqrt{-\bar g}$. Taking into account Eq. (3.2) we complete the demonstration that (2.5) is invariant in form under the transformation of units (3.1), together with the scalar field redefinition (3.2) and the free-parameter transformation (3.4) with $\sigma\neq 1$ even when matter fields are present. 

In view of the partial results of this section, it should be remarked that, although the laws of gravity that are derivable from the action (2.5), in general, are not invariant under (3.1), (3.2) when $\sigma$ is arbitrary [this includes the case when $\sigma=1$], they are invariant under these transformations [plus (2.14) and (2.18)] when $\sigma\neq 1$, i. e., they are invariant under the transformations of the "group of units transformations". Hence, we reach to the conclusion that, in view of the requirement of the Brans-Dicke postulate, the conformal formulation of general relativity (or string frame formulation) is a consistent formulation of the laws of gravity and, therefore, should be taken as a serious alternative for the formulation of the laws of gravity. Consequently, observational evidence should be geometrically interpreted on the grounds of Weyl-integrable geometry.

\section{Weyl geometry and spacetime singularities}

Yet another consequence is related with the conformal transformations of the kind (1.1). It is connected with the spacetime singularities that usually arise in Riemannian manifolds. Under (1.1) the Ricci tensors with a hat and without it are related through

\begin{equation}
R_{ab}=\hat R_{ab}-3\hat g_{ab}\Omega^{-2}(\hat\nabla\Omega)^2+\Omega^{-1}(2\Omega_{;ab}+\hat g_{ab}\hat\Box\Omega),
\end{equation}
where the covariant derivative in the right hand side(RHS) of eq.(2.11) is given in respect to the metric ${\hat g_{ab}}$. Hence the condition $R_{mn}K^m K^n\geq 0$ is transformed into the following condition:

\begin{equation}
\hat R_{mn}\hat K^m \hat K^n-3\hat g_{mn}\hat K^m \hat K^n\Omega^{-2}(\hat\nabla\Omega)^2+2\Omega^{-1}\Omega_{;mn}\hat K^m \hat K^n+\hat g_{mn}\hat K^m \hat K^n\Omega^{-1}\hat\Box\Omega\geq 0,
\end{equation}
where the non-spacelike vector $\hat K^a (\hat g_{mn}\hat K^m \hat K^n\leq 0)$ is related with $K^a$ through $\hat K^a=\Omega^{-1}K^a$. This means that the condition (2.12) may be fulfilled even if $\hat R_{mn}\hat K^m \hat K^n<0$ and, correspondingly, in the conformal manifold with metric $\hat g_{ab}$ some singularity theorems may not hold. Therefore, in principle, spacetime singularities occurring in Riemann geometry may be removed in its equivalent Weyl-integrable geometry generated by the physically  meaningless transformation of units (1.1). It is very encouraging since, as shown in previous sections, Weyl-integrable geometry meets the requirements of BD postulate and, consequently, represents a consistent geometrical setting where to interpret the observational evidence.

Without loss of generality we shall study the Raychaudhuri equation for a congruence of time-like geodesics without vorticity, with the time-like tangent vector $V^a$:

\begin{equation}
\frac{d\Theta}{ds}=-R_{mn}V^m V^n-2\sigma^2-\frac{1}{3}\Theta^2,
\end{equation}
where $\Theta$ is the volume expansion of the time-like geodesic and $\sigma$ is the shear. As seen from (4.1) $\Theta$ will monotonically decrease along the time-like geodesic if $R_{mn}K^m K^n\geq 0$ for any time-like vector $\bf K$(the time-like convergence condition).

Under (1.1) the Raychaudhuri equation (4.1) is mapped into:

\begin{equation}
\frac{d\hat\Theta}{d\hat s}=-\hat R_{mn}\hat V^m \hat V^n-2\hat\sigma^2-\frac{1}{3}\hat\Theta^2+\Omega^{-1}\Omega_{,n}\hat V^n \hat\Theta+\Omega^{-1}(\Omega_{;nm}-3\frac{\Omega_{,n}\Omega_{,m}}{\Omega})\hat h^{nm},
\end{equation}
where $\hat h^{ab}=\hat g^{ab}+\hat V^a \hat V^b$. The additional fourth and fifth terms in the right-hand side(RHS) of Eq. (4.2) have non-definite sign and hence can, in principle, contribute to expansion instead of contraction. When the contribution to expansion of these terms (if they effectively contribute to expansion) becomes stronger than the contribution to focusing of the first three terms, then contraction changes into expansion and no singularity occurs. In this case wormhole spacetimes are allowed instead of singular ones. A less ambiguous discussion of this subject could be given only after incorporation of an effective theory of spacetime.

As an effective theory of spacetime we shall approach general relativity with an extra scalar field $\psi$, given by the canonical action Eq. (2.4). The equations derivable from this action are

\begin{equation}
G_{ab}=8\pi T_{ab}+\alpha(\psi_{,a}\psi_{,b}-\frac{1}{2} g_{ab}(\nabla\psi)^2),
\end{equation}
where the gravitational constant $G=1$, $G_{ab}\equiv R_{ab}-\frac{1}{2}g_{ab}R$, $T_{ab}=\frac{2}{\sqrt{-g}}\frac{\partial (\sqrt{-g}L_{matter})}{\partial g^{ab}}$ is the stress-energy tensor for the matter fields, and $\frac{1}{8\pi}$ times the 2nd term in the rhs of Eq.(4.5) is the stress-energy tensor for the scalar field. The following wave equation for $\psi$ is also derivable from the action (2.4),

\begin{equation}
\Box\psi=0.
\end{equation}

The stress-energy tensor $T_{ab}$ fulfills the conservation equation

\begin{equation}
T^{an}_{;n}=0.
\end{equation} 

For illustration we shall study two generic situations. 

\subsection{The Schwarzschild black hole}

In this section, for simplicity, we shall interested in the static, spherically symmetric solution to Riemannian general relativity (Einstein frame GR with $\alpha\geq 0$) for material vacuum, and in its conformal Weyl-integrable picture (conformal GR). In this case the field Eq. (4.5) can be simplified

\begin{equation}
R_{ab}=\alpha\psi_{,a}\psi_{,b}.
\end{equation}

We shall study separatelly the cases with $\alpha=0$ and $\alpha>0$. For $\alpha=0$ the solution to Eq. (4.8) [in Schwarzschild coordinates] looks like 

\begin{equation}
ds^2=-(1-\frac{2m}{r}) dt^2+ (1-\frac{2m}{r})^{-1} dr^2+ r^2 d\Omega^2,
\end{equation}
where $d\Omega^2=d\theta^2+\sin^2 \theta d\varphi^2$. Meanwhile the solution to Eq. (4.6) is found to be

\begin{equation}
\psi=q \ln (1-\frac{2m}{r}),
\end{equation}
where $m$ is the mass of the point, located at the coordinate beginning, that generates the gravitational field and $q$ is an arbitrary real parameter. As seen from Eq.(4.9) the static, spherically symmetric solution to Eq.(4.8) is just the typical Schwarzschild black hole solution for vacuum. The corresponding solution for conformal GR can be found with the help of the conformal rescaling of the metric (1.1) with $\Omega^2=\phi=e^\psi=(1-\frac{2m}{r})^q$:

\begin{equation}
d\hat s^2=-(1-\frac{2m}{r})^{1-q} dt^2+(1-\frac{2m}{r})^{-1-q} dr^2+\hat\rho^2d\Omega^2,
\end{equation}
where we have defined the proper radial coordinate $\hat\rho=r(1-\frac{2m}{r})^{-\frac{q}{2}}$. In this case the curvature scalar given in terms of the metric with a hat is given by 

\begin{equation}
\hat R=-\frac{3}{2} \hat g^{nm} \hat\nabla_n \psi \hat\nabla_m \psi=-\frac{6 m^2 q^2}{r^4}(1-\frac{2m}{r})^{q-1}.
\end{equation}

The real parameter $q$ labels different spacetimes ($M,\hat g^{(q)}_{ab},\psi^{(q)}$), so we obtained a class of spacetimes \{$(M,\hat g^{(q)}_{ab},\psi^{(q)})/q\in \Re$\} that belong to a bigger class of known solutions\cite{alac}. These known solutions are given, however, for an arbitrary value of the coupling constant $\omega=\alpha-\frac{3}{2}$.

We shall outline the more relevant features of the solution given by (4.6). For the range $-\infty<q<1$ the Ricci curvature scalar (4.7) shows a curvature singularity at $r=2m$. For $-\infty<q<0$ this represents a timelike, naked singularity at the origin of the proper radial coordinate $\rho=0$. Situation with $q=0$ is trivial. In this case the conformal transformation (1.1) coincides with the identity transformation that leaves the theory in the same frame. For $q>0$, the limiting surface $r=2m$ has the topology of an spatial infinity so, in this case, we obtain a class of spacetimes with two asymptotic spatial infinities -one at $r=\infty$ and the other at $r=2m$- joined by a wormhole with a throat radius $r= (2+q)m$, or the invariant surface determined by $\hat\rho_{min} =q(1+\frac{2}{q})^{1+\frac{q}{2}} m$. 

This way, Weyl-integrable spacetimes conformal to the Riemannian Schwarzschild black hole one are given by the class \{$(M,\hat g^{(q)}_{ab},\psi^{(q)})/q>0$\} of wormhole (singularity free) spacetimes. 

When $\alpha>0$ there is a physical scalar in the Einstein frame. The corresponding solution to Eq.(4.8) is given by\cite{alac}: 

\begin{equation}
ds^2=-(1-\frac{2m}{pr})^p dt^2+ (1-\frac{2m}{pr})^{-p} dr^2+\rho^2 d\Omega^2,
\end{equation}
while the solution to Eq. (4.6) is found to be

\begin{equation}
\psi=q \ln (1-\frac{2m}{pr}),
\end{equation}
where $p^2+2\alpha q^2=1$, $p>0$. For non-exotic scalar matter [$\alpha \geq 0$] $0<p\leq 1$. In Eq. (4.13) we have used the definition $\rho=(1-\frac{2m}{pr})^\frac{1-p}{2}r$ for the Einstein frame proper radial coordinate. There is a time-like curvature singularity at $r=\frac{2m}{p}$, so the horizon is shrunk to a point. Then in the Einstein frame the validity of the cosmic censorship hypothesis and, correspondingly, the ocurrence of a black hole are uncertain\cite{alac}.

The solution conformal to (4.13) is given by

\begin{equation}
d\hat s^2=-(1-\frac{2m}{pr})^{p-q} dt^2+ (1-\frac{2m}{pr})^{-p-q} dr^2+ \hat \rho^2 d\Omega^2,
\end{equation}
where the conformal proper radial coordinate $\hat\rho=r(1-\frac{2m}{pr})^{\frac{1-p-q}{2}}$ was used. In this case, when $\alpha$ is in the range $0 < \alpha+\frac{3}{2} < \frac{1+p}{2(1-p)}$, the Weyl-integrable spacetime shows again two asymptotic spatial infinities joined by a wormhole.

The singularity-free character of the Weyl-integrable geometry should be tested with the help of a test particle that is acted on by the conformal metric in Eq.(4.15) and by the scalar field $\psi=q \ln(1-\frac{2m}{pr})$. Consider the radial motion of a time-like test particle ($d\Omega^2=0$). In this case the time-component of the motion equation (2.10) can be integrated to give

\begin{equation}
\dot t^2=-C_1^2(1-\frac{2m}{pr})^{q-2p}, 
\end{equation}
where $C_1^2$ is some integration constant and the overdot means derivative with respect to the conformal proper time $\tau$ ($d\tau^2=-d\hat s^2$). The integration constant can be obtained with the help of the following initial conditions: $r(0)=r_0$, $\dot r(0)=0$, meaning that the test particle moves from rest at $r=r_0$. We obtain $C_1^2=-(1-\frac{2m}{pr_0})^p$. Then the proper time taken for the particle to go from $r=r_0$ to the point with the Schwarzschild radial coordinate $r_0\leq r < \frac{2m}{p}$ is given by

\begin{equation}
\tau=\int_r^{r_0}\frac{r^{\frac{q}{2}} dr}{\sqrt{(1-\frac{2m}{pr_0})^p-(1-\frac{2m}{pr})^p}(r-\frac{2m}{p})}.  
\end{equation}

While deriving this equation we have used Eq. (4.15) written as follows: $-1=(1-\frac{2m}{pr})^{p-q} d\dot t^2- (1-\frac{2m}{pr})^{-p-q} d\dot r^2$. The integral in the RHS of Eq. (4.15) can be evaluated to obtain ($q\neq 2$)

\begin{equation}
\tau>\frac{(\frac{2m}{p})^{\frac{q}{2}}(1-\frac{2m}{pr_0})^{\frac{p}{2}}}{1-\frac{q}{2}}[(r_0-\frac{2m}{p})^{1-\frac{q}{2}}-(r-\frac{2m}{p})^{1-\frac{q}{2}}], 
\end{equation}
and

\begin{equation}
\tau>\frac{2m}{p} \ln(\frac{r_0-\frac{2m}{p}}{r-\frac{2m}{p}}),  
\end{equation}
for $q=2$. For $q\geq 2$ the proper time taken by the test particle to go from $r=r_0$ to $r=\frac{2m}{p}$ is infinite showing that the particle can never reach this surface (the second spatial infinity of the wormhole). Then the time-like test particle does not see any singularity in the spacetime of a WIST configuration that is given by the line-element (4.15).

If we consider the scalar field $\psi$ as a perfect fluid then we find that its energy density as measured by a comoving observer is given [in terms of WIST geometry] by\cite{iq}

\begin{equation}
\hat\mu^\psi=\frac{2m^2 q^2 \alpha}{8\pi p^2 r^4}(1-\frac{2m}{pr})^{p+2q-2}.  
\end{equation}

It is everywhere non-singular for $q\geq\frac{2-p}{2}$ ($0<p\leq 1$) in the range $2m\leq r<\infty$. This means that the scalar matter is non-exhotic and shows a non-singular behaviour evrywhere in the given range of the parameters involved.  
 
Although the wormhole picture is not simpler than its conformal black hole one, it is more viable because these geometrical objects (WIST wormholes) are invariant respecting the transformations of the units of length, time and mass as it was shown in the former section. As noted by Dicke\cite{dk}, these transformations should not influence the physics if the theory is correct. The Riemannian Schwarzschild black hole, for his part, does not possess this invariance.

\subsection{Friedmann-Robertson-Walker cosmology}

Now we shall study homogeneous and isotropic universes with the Friedmann-Robertson-Walker(FRW) line-element [given in terms of the Riemann metric $g_{ab}$]

\begin{equation}
ds^2=-dt^2+a^2(t)(\frac{dr^2}{1-kr^2}+r^2d\Omega^2),
\end{equation}
where $a(t)$ is the scale factor, $d\Omega^2=d\vartheta^2+\sin^2\vartheta d\varphi^2$ and $k=0$ for flat, $k=-1$ for open and $k=+1$ for closed universes. The universe is supposed to be filled with a barotropic perfect fluid with the barotropic equation of state $p=(\gamma-1)\mu$, where $\mu$ is the energy density of matter and the barotropic index $0<\gamma<2$. The perfect fluid stress-energy tensor is $T_{ab}=(\mu+p)V_a V_b+p g_{ab}$. Equations (4.5), (4.6) and (4.7) lead to the following equation for determining the scale factor

\begin{equation}
(\frac{\dot a}{a})^2+\frac{k}{a^2}=\frac{M}{a^{3\gamma}}+\frac{\alpha A^2}{a^6},
\end{equation}
where the overdot means derivative with respect to the proper time $t$. $M$ and $A$ are arbitrary integration constants. While deriving Eq. (4.22) we have considered that after integrating Eq. (4.6) once we obtain $\dot\psi=\pm\frac{\sqrt{6}A}{a^3}$, while (4.7) gives $\mu=\frac{3M}{8\pi a^{3\gamma}}$.

If we take the time-like tangent vector of a comoving observer $V^a=\delta^a_0$, then the Raychaudhuri equation (4.3) can be written as

\begin{equation}
\dot\Theta^2=-\frac{9}{2}\frac{\gamma M}{a^{3\gamma}}-\frac{9\alpha A^2}{a^6}+\frac{3k}{a^2},
\end{equation}
where we took the reversed sense of the proper time $-\infty\leq t\leq 0$, i. e., $a(t)$ runs from infinity to zero. The well-known results of GR emerge from a careful analysis of this equation. For $k=0$ and $k=-1$, Eq. (4.23) gives focusing of the fluid lines, leading to a global singularity at $t=0$ where the density of matter becomes infinite and the known physical laws breakdown. For $k=+1$ $a(t)$ is a monotonic function of the proper time. It grows from zero to a maximum value and then decreases to zero. Eq. (4.23) gives focusing of the fluid lines twice (the value $a=0$ occurs twice) as required. The closed universe emerges from a global initial singularity, grows and then merges into a final singularity. All these results are well-known in general relativity on Riemann manifolds. Our goal here is to show what happens when we formulate general relativity on a manifold of WIST structure.

If we set $\Omega$ in Eq.(1.1) to be $\Omega^2=e^\psi$ then the Raychaudhuri equation in the corresponding Weyl-integrable manifold can be written in terms of the Riemann scale factor $a(t)$ as

\begin{equation}
\frac{d\hat\Theta^\pm}{d\tau}=e^{-\psi^\pm}(-\frac{9}{2}\frac{\gamma M}{a^{3\gamma}}-\frac{9(\alpha+\frac{1}{2}) A^2}{a^6}+\frac{3k}{a^2}\pm \frac{6\sqrt{6}A}{a^3}\sqrt{\frac{\gamma M}{a^{3\gamma}}+\frac{\alpha A^2}{a^6}-\frac{k}{a^2}}),
\end{equation}
where $\tau=\int e^\frac{\psi}{2} dt$ is the proper time in the Weyl spacetime generated by the transformation (1.1) with $\Omega=e^\frac{\psi}{2}$ applied to the FRW -Riemannian- spacetime with the line-element (4.21). For simplicity we shall concentrate on flat ($k=0$) and open ($k=-1$) universes (the case $k=+1$ requires a very detailed analysis). For big $a$ Eq. (4.24) shows that focusing of the fluid lines occurs. As we go backwards in time $t$, $a$ decreases and, for sufficiently small $a\ll 1$, Eq. (4.24) can be written as

\begin{equation}
\frac{d\hat\Theta^\pm}{d\tau}\approx \frac{3A^2e^{-\psi^\pm}}{a^6}(-3\alpha\pm 2\sqrt{6}\sqrt{\alpha}-\frac{3}{2}).
\end{equation}

If we choose the '+' branch of the solution of the wave equation (4.6)

\begin{equation}
\psi^\pm=\psi_0\mp\sqrt{6A}\int\frac{da}{a^3 \dot a},
\end{equation}
that is given by the choice of the '+' sign in (4.26) ($\psi_0$ is an integration constant), then when $\alpha$ is in the range $0\leq\alpha\leq\frac{1}{6}$ and for small enough $a$, Eq.(4.24) [and correspondingly Eq. (4.25)] shows that contraction of fluid lines turns into expansion and no singularity is formed in the Weyl-integrable spacetime conformal to the Riemannian FRW one. In fact, while the curvature scalar $R$ in terms of the original [Riemannian] metric

\begin{equation}
R=\frac{3}{a^6}[(4-3\gamma)M a^{3(2-\gamma)}-2\alpha A^2],
\end{equation}
is singular at $a=0$, its conformal (in the '+' branch and for small $a$) behaves like

\begin{equation}
\hat R^+\sim -3(2\alpha-3)A^2 e^{\phi_0} a^{\sqrt{\frac{6}{\alpha}}-6}.
\end{equation}

For $\alpha\leq\frac{1}{6}$, $\hat R^+$ is bounded even for $a=0$. Moreover, for $k=0$ and $k=-1$, and $0\leq\alpha\leq\frac{1}{6}$, $\hat R^+$ is regular and bounded everywhere in the range $0\leq a\leq\infty$. In the spacetime of WIST configuration the fluid lines converge into the past up to the moment when $a$ becomes small enough and then they diverge. This is the way the spacetime singularity that always occurs in the FRW [Riemannian] spacetime given by (4.21), is removed in its conformally-Riemannian spacetime.

The following question is to be raised. Is illusory the vanishing of the cosmological singularity in the spacetime of WIST configuration?. In other words: would fluid particles feel a singularity in the Weyl-integrable spacetime even if the geometry seems to be that of a cosmological wormhole?. In this sense we shall remark that the inevitabillity of the cosmological singularity in the Riemann manifold with the FRW line-element (4.21) is linked with the fact that the fluid lines are incomplete into the past [the proper time $t$ is constrained to the range $0\leq t\leq +\infty$]. In its conformal wormhole spacetime given by the line-element

\begin{equation}
d\hat s^2=-d\tau^2+\hat a^2(t)(\frac{dr^2}{1-kr^2}+r^2d\Omega^2),
\end{equation}
where $\hat a=e^{\frac{\psi}{2}}a$, the fluid lines are complete in the manifold of WIST structure. In particular, when $t$ goes over the range $0\leq t\leq +\infty$, $\tau$ goes over the range $-\infty\leq\tau\leq +\infty$. While the matter density seen by a comoving observer in the Riemann geometry $T_{mn}V^m V^n=\mu$ ($V^a=\delta^a_0$) is singular at $t=0$($a=0$), the corresponding energy density as measured by a comoving observer in Weyl-integrable geometry ($\hat V^a=e^{\frac{\psi^+}{2}}V^a$), $\hat T_{mn}\hat V^m \hat V^n=e^{2\psi^+}\mu$ is regular and bounded for all times $0\leq t\leq +\infty$ ($-\infty\leq\tau\leq +\infty$). The same is true for the energy density of the scalar field $\hat\mu_\psi$ measured by a comoving observer. Hence, the absence of spacetime singularities in a Weyl-integrable spacetime (conformal to a singular spacetime of Riemann structure) is a real feature of this geometry.

We should explain yet another thing. The vanishing of the cosmological singularity in the Weyl spacetime is allowed only in the '+' branch of our solution. In the '-' branch the Weyl spacetime is singular too (like the Riemann one). Hence we should give a physical consideration why we chose the '+' branch. In this sense we shall note that under the conformal transformation (1.1) with $\Omega^2=e^\psi$, the action (2.4) is mapped into its conformal action (2.5).

Hence in the Weyl-integrable manifold $e^{-\psi}$ plays the role of an effective gravitational constant $\hat G$. For the '-' branch $\hat G$ runs from zero to an infinite value, i.e., gravity becomes stronger with the evolution of the universe and in the infinite future it dominates over the other interactions (or becomes of the same range), that is in contradiction with the usual unification scheme. On the contrary, for the '+' branch, $\hat G$ runs from an infinite value to a finite constant value as the universe evolves and, hence, gravitational effects are weakened as required. The fact that, in this branch, the vanishing of the singularity is effective only for $0\leq\alpha\leq\frac{1}{6}$ can be taken only as a restriction on the values the free parameter $\alpha$ can take.

Exact analytic solutions for conformal general relativity with a barotropic perfect fluid can be found in Ref. \cite{qbc} for flat FRW cosmology and in Ref. \cite{qcb} for open dust-filled and radiation-filled universes.

Finally I shall remark that, since conformal general relativity is invariant under the transformations of the units of length, time and mass, the WIST wormhole solution (Eq. (4.11) for $\alpha=0$ and Eq. (4.15) for $\alpha>0$) and the FRW bouncing spacetime of WIST structure Eq. (4.29) -being obtained within the context of conformal GR- are invariant objects under (3.1), (3.2) and (3.4) with $\sigma\neq 1$. Their conformal singular [Riemannian] spacetimes do not share this symmetry. Hence, following symmetry arguments, the picture without singularity is more viable than the one with spacetime singularities.

\section{Weyl geometry and the quantum}

There are some results that hint at Weyl-integrable geometry as a geometry that can take account, in a natural way, of the quantum effects of matter. In Ref. \cite{novello,rankin}, for instance, the authors suggested a deep connection between quantum mechanics and Weyl structures. This connection had been made evident in the early times of London\cite{london}. He found that some quantum rules could be obtained from a classical description based on Weyl geometry. Another argument may be found in Ref. \cite{dzhunu} where the authors demonstrated that there is some kind of correlation between objects in classical gravity on Weyl manifolds and in quantum non Abelian field theory.

In the present section I should like to present other arguments in this direction. The first but, not the unique, is linked with the following circunstance. Weyl-integrable geometry is already invariant under the group of units transformations [Eq. (3.1), (3.2) and (3.4) with $\sigma\neq 1$] as required for any geometrical setting where to describe the physics that is itself insensible to the transformations of the units of measure one chooses. I hope that a final quantum theory of spacetime should reflect this invariant behaviour.

Another interesting argument in this sense is connected with an idea presented in reference \cite{shojai}. These papers were based on the de Broglie-Bohm quantum theory of motion\cite{bohm}. The authors showed that the quantum effects of matter, being explicit in one frame through the following expression,  $(\nabla S)^2=m^2(1+Q)$, where $S$ is the canonical action for the matter fields, $m$ is the constant mass of the matter particles and $Q$ is the matter quantum potential, can be hidden in a conformal transformation of the kind (1.1) in the conformal formulation of this expression. They concluded that the quantum effects of matter are already contained in the conformal metric they called as "physical metric". They were led to this conclusion since, the non-geodesic motion of matter particles in one frame (due to the quantum force), is mapped into (apparently) a geodesic motion in the conformal frame if one considers that in (1.1) $\Omega^2=1+Q$. We are not concerned here with the validity of these results. Our approach is a little different although the leading idea is the same. We shall take the action (2.4) for GR in the Einstein's formulation that does not contain the quantum effects. In this formulation material test particles follow the geodesics of the Riemann geometry

\begin{equation}
\frac{d^2 x^a}{ds^2}+\{\;\;^a_{mn}\}\frac{dx^m}{ds}\frac{dx^n}{ds}=0.
\end{equation}

Under the conformal rescaling (1.1) with $\Omega^2=e^{\psi}$, this equation is mapped into

\begin{equation}
\frac{d^2 x^a}{d\hat s^2}+\{\;\;^a_{mn}\}_{hat}\frac{dx^m}{d\hat s}\frac{dx^n}{d\hat s}=\frac{1}{2}\psi_{,n}(\frac{dx^n}{d\hat s}\frac{dx^a}{d\hat s}-\hat g^{an}).
\end{equation}

If we set $e^{\psi}=1+Q$, where $Q$ is the matter quantum potential, hence we can consider that the RHS of Eq. (5.2) is the quantum force. This means that under the conformal rescaling, the classical motion given by (5.1) is mapped into a quantum motion in the conformal frame. In fact, under a conformal transformation of
the kind (1.1) with $\Omega^2=e^\psi=1+Q$, the geodesic equations of Riemann geometry [Eq.(5.1)] are
mapped into the Eq.(5.2) defining a non-geodesic motion on a spacetime with metric $\hat g_{ab}$ provided that the Riemaniann structure of the geometry is preserved under (1.1).

However, we must acknowledge that Eq. (5.2) defines, in fact, a time-like geodesic in a Weyl-integrable manifold (as discussed in Sec. II of this paper). This means that a free-falling test particle would not "feel" the quantum force if the motion is interpreted on the grounds of a WIST geometry and, correspondingly, the metric with a hat $\hat g_{ab}$ is taken to give the physical interpretantion of the experimental observations. This hints at the conclusion that Weyl-integrable geometry contains implicitly the quantum effects of matter, i.e., it is already a quantum geometry\cite{bqc}.

Yet another result that hints at Weyl geometry as a geometry that implicitly contains the quantum effects of matter has been already presented in Sec. IV. In the usual Einstein's formulation of general relativity, the occurrence of spacetime singularities is inevitable if the matter obeys some reasonable energy conditions. It is usually linked with the lack of quantum considerations in this formulation of GR. This can be thought of as a property of Riemann spacetimes in general. Hence, we can regard Riemann geometry as an "incomplete" geometry since it must be "completed" with the inclusion of quantum effects. In general, we are tempted to call geometries compatible with geodesically incomplete spacetimes as incomplete geometries. Then we are led to consider the incompleteness of a given geometry and, correspondingly the ocurrence of spacetime singularities in it, as due to the fact that it does not consider the quantum effects of matter.

As we have already shown in Sec. III for both Schwarszchild spacetime and FRW cosmology, singularities occurring in usual Einstein's GR that is naturally linked with Riemann geometry, are removed in the conformal formulation that is naturally linked with Weyl-integrable geometry. Singular Riemannian spacetimes are mapped under the conformal rescaling (1.1) into wormhole [singularity-free] spacetimes of WIST structure. In other words, in these cases the incomplete Riemann geometry is conformal to a complete Weyl-integrable geometry. We hope that completeness of Weyl-integrable geometry means that it implicitly contains the quantum effects of matter.

If our considerations here are correct then, we can reach to the following conclusion. Complete Weyl-integrable geometries should be taken as a proper framework for describing the physical laws of nature without the unnatural separation of physics in classical and quantum laws. Consequently, conformal (string frame) GR in which the underlying manifold is of WIST configuration, provides an intrinsically quantum description of gravity.

\section{Is the geometry of the world unique?}

The fact that physical observations can not distinguish a Riemann manifold with singularities from a Weyl-integrable manifold without them, is very striking. Experimental observations show, in particular, that there are several astrophysical black holes located in our universe. One of them is located in the center of our own galaxy.

In this paper [and for the first time in Ref. \cite{iq}] it has been shown that under a conformal rescaling of the kind (1.1), what appears as a Schwarzschild black hole in a Riemann spacetime is mapped into a wormhole (singularity-free) spacetime of WIST configuration. Since the conformal transformation (1.1) does not touch the spacetime coincidences [coordinates], i. e., spacetime measurements are unchanged under this transformation, hence we can conclude that a Riemannian Schwarzschild black hole is observationally indistinguishable from a Weyl-integrable wormhole spacetime. Although an astrophysical black hole does not has the high symmetry inherent to a Schwarzschild black hole, we expect that under (1.1) it is transformed into a some kind of astrophysical wormhole or a similar astrophysical object without singularity. Hence we arrive at a kind of "duality" of the geometrical representation of our real [observationally testable] world. Our previous discussion resolves this "duality". In effect, Riemann geometry [with the undesirable occurrence of singularities] is not a consistent framework for confronting our experimental observations since it is not invariant under a general transformation of units (in particular the transformation of units studied in the present paper), while the experimental measurements should not depend on the particular values of the units of measure one chooses. In this context, we may interpret the occurrence of black holes (in particular astrophysical black holes when experimental observations are concerned) as due to a wrong choice of the geometrical framework for describing the laws of gravity [and, consequently, as due to a wrong choice of the formulation of the laws of gravity]. Weyl-integrable geometry seems to be this correct geometrical framework since the basic requirement under which it rests [Eq. (2.6)] is not affected by the units transformation studied here. Hence, although the occurrence of compact astrophysical objects has been made evident by the experimental observations, these should not be confronted as Riemannian black holes containing singularities but as wormholes of Weyl-integrable configuration without them.

A final remark about the transformations (3.1), (3.2) with $\sigma\neq 1$. These can be viewed as a transformation in the space of the parameter of the theory $\alpha$. In effect, under the above transformations we move within the parameter space $\alpha$ so the set of spacetimes $\{({\cal M},\hat g^{(\alpha)}_{ab},\psi^{(\alpha)})/\alpha\in\Re\}$, where $\cal M$ is a smooth manifold of Weyl-integrable configuration, forms an equivalence class. The spacetimes in this class are physically equivalent for the representation of the physical reality. This equivalence includes, of course, observational equivalence. The physical laws look the same in all of the spacetimes that belong to the former equivalence class. We are then led to a kind of postulate of equivalence of geometries: {\it There exists an infinite set of spacetimes $({\cal M},\hat g^{(\alpha)}_{ab})$  of WIST configuration, that are physically equivalent and equally consistent, for the description of the physical laws}. This can be viewed as an extension of the postulate of equivalence of coordinate systems linked by general coordinate transformations in GR.

The consequence of the postulate of equivalence of conformally-Riemannian  [Weyl-integrable] geometries for the description of the physical reality I shall call as "geometrical relativity". I think this will represent a natural extension of general relativity to include invariance under the one-parameter group of point-dependent transformations of the units of lenght, time and mass and, if our considerations in Sec. V are correct, to implicitly include the quantum effects of matter.

Finally we shall remark that the relativity of geometry implies nothing but just relativity of the geometrical interpretation of the physical reality. Physical reality itself is unique.

It will be of interest, in the future, to look at more general transformations of units than those given in this paper in order to further extend our results.

\begin{center}
{\bf ACKNOWLEDGEMENT}
\end{center}

We acknowledge useful discussions with colleagues Rolando Bonal and Rolando Cardenas and MES of Cuba by financial support.

\end{document}